\documentclass[aps,prl,twocolumn,groupedaddress,showpacs]{revtex4-1}
\usepackage{graphicx}  
\usepackage{dcolumn}   
\usepackage{bm}        
\usepackage{amssymb,amsfonts,amsmath}   
\usepackage{color}
\usepackage{lipsum}
\usepackage{upgreek}

\usepackage{float}
\newfloat{vidfigure}{placement}{ext}
\floatname{vidfigure}{Video}

\usepackage[normalem]{ulem}

\renewcommand{\d}{\mathrm{d}}

\renewcommand{\d}{\mathrm{d}}
\renewcommand{\v}{{\boldsymbol v}}
\newcommand{\x}{{\boldsymbol x}}

\allowdisplaybreaks

\begin{document}

\title{Aerotaxis in the Closest Relatives of Animals}

\author{Julius B. Kirkegaard}
\author{Ambre Bouillant}
\author{Alan O. Marron}
\author{Kyriacos C. Leptos}
\author{Raymond E. Goldstein}
\affiliation{Department of Applied Mathematics and Theoretical Physics, Centre for Mathematical Sciences, \\ University of 
Cambridge, Wilberforce Road, Cambridge CB3 0WA, United Kingdom}

\date{\today}

\begin{abstract}
As the closest unicellular relatives of animals, choanoflagellates serve as useful model 
organisms for understanding the evolution of animal multicellularity.
An important factor in animal evolution was the increasing ocean oxygen levels in the Precambrian, which are thought to have 
influenced the emergence of complex multicellular life.  As a first step in 
addressing these conditions, we study here the response of the colony-forming choanoflagellate \textit{Salpingoeca rosetta} to oxygen gradients.
Using a microfluidic device that allows spatio-temporal variations in oxygen concentrations,
we report the discovery that \textit{S. rosetta} display positive aerotaxis. 
Analysis of the spatial population distributions provides evidence for logarithmic sensing of oxygen, which enhances sensing in 
low oxygen neighborhoods.
Analysis of search strategy models on the experimental colony trajectories finds that choanoflagellate aerotaxis is consistent with 
stochastic navigation, the statistics of which are captured using an effective continuous version based on classical run-and-tumble chemotaxis.
\end{abstract}

\maketitle

\section{Introduction}
\textit{Taxis}, the physical migration towards preferred or away from undesired conditions, is a feature shared by virtually all 
motile organisms. Taxis comes in many forms, and in common is an underlying field of attractant (or repellent) and an ability to react and navigate along gradients of this field.
For example, bacteria do \textit{chemotaxis} towards nutrients \cite{Adler1969,Berg1993} and away from toxins \cite{Tso1974}.
Algae do \textit{phototaxis} towards light \cite{Yoshimura2001,Drescher2010} and \textit{gyrotaxis} along gravitational potentials \cite{Kessler1985}.
Chemotaxis provides a mechanism for the recognition and attraction of gametes \cite{Vogel1982} and for complex behavioural patterns 
such as in the slime mould \textit{Dictyostelium discoideum}, where cAMP-driven chemotaxis is a critical part of the formation of the multicellular stage of the life cycle \cite{Bonner1947}. 
Aerotaxis, defined as oxygen-dependent migration, is well-characterized in bacteria \cite{Taylor1999},
but is poorly studied in more complex organisms.
This is despite the essentiality of oxygen for all aerobic life, and the important role that
Precambrian oxygen levels played in the emergence and evolution of multicellular animal life \cite{Nursall1959}.

One group of aquatic heterotrophic protists, the choanoflagellates, are of particular interest for the study of how multicellularity evolved. Choanoflagellates 
are a class of unicellular microorganisms that are the closest relatives of the animals \cite{Lang2002}. This relationship was first proposed by James-Clark in 1866 \cite{James-Clark1866}, on the basis of the resemblance between choanoflagellates and the choanocytes of sponges. 
The sister relationship between choanoflagellates and animals was further confirmed in the genomic era by molecular evidence \cite{King2008}.
All choanoflagellates share the same basic unicell structure: a prolate cell body with a 
single beating flagellum that is surrounded by a collar of microvilli. The beating of the flagellum
creates a current in the surrounding fluid that guides suspended prey such as bacteria through the collar \cite{Pettitt2002} where they can be caught
and ultimately phagocytosed. The flagellar current also has the effect of causing the choanoflagellate
cell to swim.

\begin{figure}[b]
\centering
\includegraphics[width=0.98\linewidth]{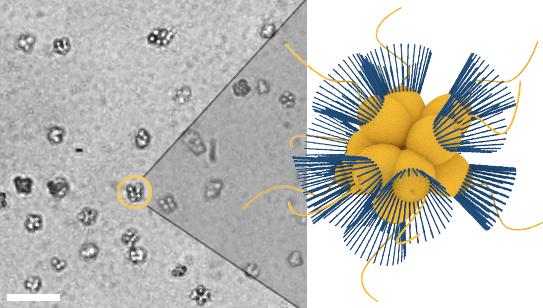}
\caption{Micrograph of \textit{S. rosetta} colonies (left) with schematic illustration (right, collars in blue).
Scale bar: 50 $\upmu$m. Cell body diameters are $\sim 5 \, \upmu$m. \label{fig:colony}}
\end{figure}

The choanoflagellate \textit{Salpingoeca rosetta} can form colonies through incomplete cytokinesis \cite{Fairclough}. In the presence of certain bacteria \cite{Dayel2011,King_eLife}, these colonies have an eponymous rosette-like shape as 
shown in Fig. \ref{fig:colony}.
The colony morphology is variable, and the constituent flagella beat independently of one another \cite{Kirkegaard2016}.
The random and independent flagellar motion argues against there being any coordination between cells in a colony, 
and as yet no evidence of any form of taxis for choanoflagellate colonies has been reported.

The geometry, cellular independence and lack of taxis observed in  \textit{S. rosetta} colonies contrast with other lineages, such as the Volvocales,
a group of green algae \cite{Goldstein2015}. 
Phototaxis is clearly observable in both unicellular (\textit{Chlamydomonas}) \cite{Yoshimura2001} and colonial (\textit{Volvox}) \cite{Drescher2010} species, in order to maintain optimum light levels for photosynthesis.
Volvocalean phototaxis is \textit{deterministic}, requiring precise tuning between the internal biochemical timescales and the rotation period of the organism as a whole.
Although \textit{S. rosetta} colonies also rotate around an internal axis,
due to the variable colony morphology and the independent beating of the individual flagella, 
this rotation rate will itself be random \cite{Kirkegaard2016},
rendering a strategy similar to that of the Volvocales unlikely in \textit{S. rosetta}.

An alternative strategy is \textit{stochastic} taxis, sometimes referred to kinesis. The classic example of stochastic taxis is the run-and-tumble chemotaxis of certain peritrichous bacteria \cite{Berg1993}.
By spinning their left-handed helical flagella in different directions, such bacteria can alternate between swimming in straight lines (running) and randomly reorienting themselves (tumbling).
Through biasing tumbles to be less frequent when going up the gradient, the bacteria exhibit biased
motion towards a chemoattractant without directly steering towards it \cite{Berg1993}.

Here, we study \textit{S. rosetta} and show that it exhibits aerotaxis, \textit{i.e.} navigation along gradients of oxygen. 
We further examine and statistically analyze aerotaxis of \textit{S. rosetta} colonies under spatio-temporal variations of oxygen at the level of total colony populations
and at the level of the trajectories of individual colonies, 
and relate the two through mathematical analysis of a generalized Keller-Segel model \cite{Keller1971} and their demonstrate the internal consistency.
From this analysis we establish two key features of the aerotactic response of choanoflagellates:
they employ a stochastic taxis search strategy and the sensing of oxygen concentration gradients is logarithmic.

\section{Results}
\subsection{Experimental Set-up}
The study of aerotaxis in bacteria has led to numerous methods for creating spatial oxygen 
gradients \cite{Shioi1987, Wong1995, Zhulin1996, Taylor1999}, one of which is the exploitation of soft lithography 
techniques \cite{Adler2012, Rusconi2014}. Since PDMS, the most commonly used material for microfluidic chambers, 
is permeable to gases, gas channels can be introduced in the devices to allow gaseous species to diffuse into the fluid.
For example, an oxygen gradient can be created using a source channel flowing with normal air and a sink channel flowing with pure nitrogen.

\begin{figure}[tb]
\centering
\includegraphics{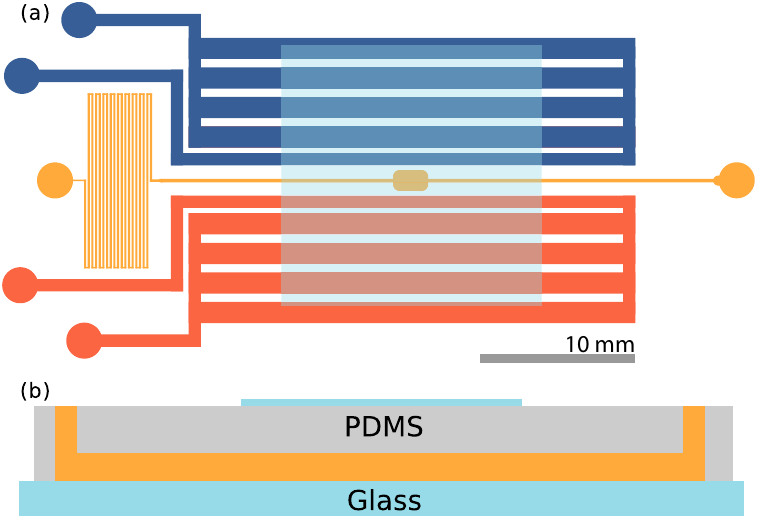}
\caption{Microfluidic device. (a) Top view of the microfluidic device. The sample channel (yellow) is loaded with culture and observed in the middle chamber. The side channels (red, blue) are gas channels in which oxygen and nitrogen may be flown. Scale bar: 10 mm. (b) Side view of the device. PDMS is plasma etched to a glass slide, and a cover slip is plasma etched on top, centered on the imaging chamber, also shown in (a). Thickness of the channels are $\approx 115 \, \upmu$m.  \label{fig:device}}
\end{figure}

Our device, shown schematically in Fig. \ref{fig:device}, is a modified version of that used by Adler, \textit{et al.} \cite{Adler2012}. 
Viewed from above, the sample channel (yellow) consists of a wide observation chamber with thin inlet and outlet channels.
The outlet leads to a serpentine channel that hinders bulk fluid flows. 
On each side of the sample channel are gas channels, the inlets of which are connected to a valve system allowing for the 
flow of air (20\% oxygen) and nitrogen.
The flow of air and nitrogen can be conducted in any combination and configuration,  \textit{e.g.} oxygen in one channel and nitrogen in the other, and can be easily swapped over.
The PDMS chamber is plasma etched to a glass slide, and an extra glass slide is etched on top of the device, preventing air from diffusing
in from the surrounding environment.

\begin{figure*}[tb]
\centering
\includegraphics[width=1\textwidth]{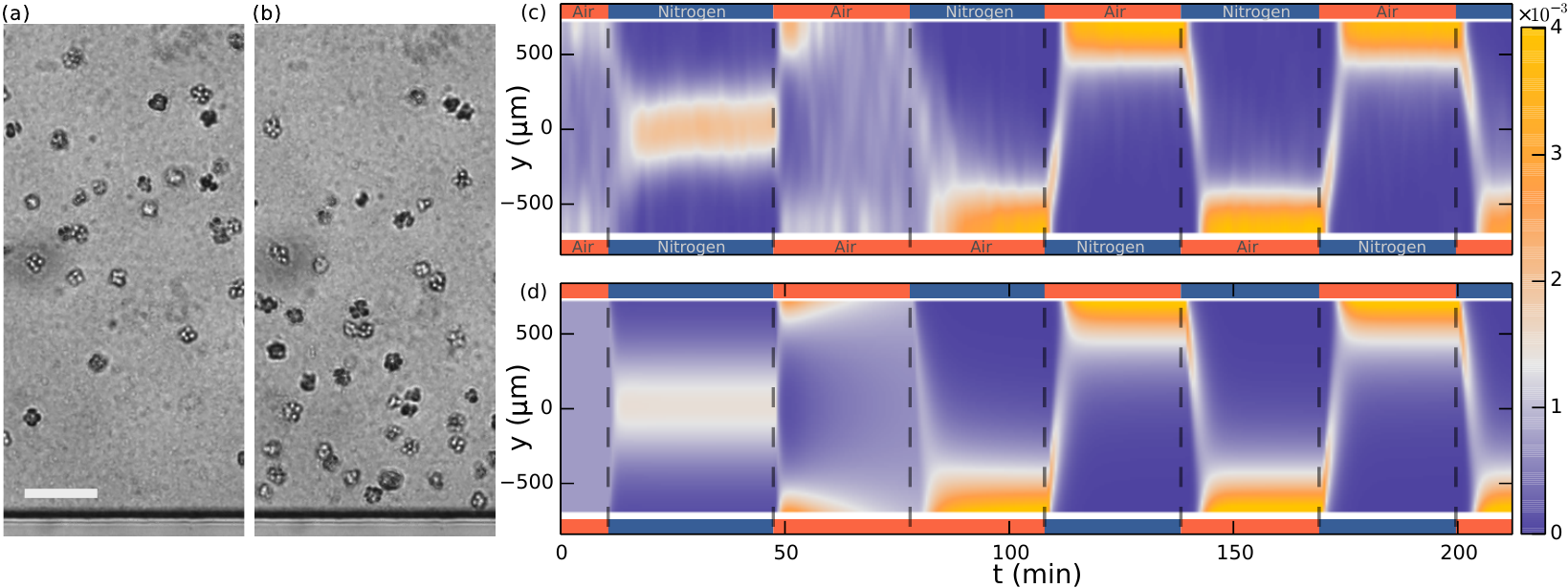}
\caption{Aerotaxis of \textit{S. rosetta} colonies. (a-b) Micrographs near an oxygen-rich wall at twice the resolution of that used in the density experiments. Scale bar: 50 $\upmu$m (a) Colonies approach a wall where the oxygen-concentration is high. (b) Colonies staying near this wall.
(c) Density evolution of \textit{S. rosetta} during experiment. At each time step the distribution is normalized to a probability distribution [colorbar units in $\upmu\text{m}^{-1}$]. Colors on the side indicate what gas is flowing in that side channel, red for oxygen and blue for nitrogen. $N_{\text{colonies}} \sim 150$, concentration $\sim 5 \cdot 10^6 \, \text{mL}^{-1}$. (d) Keller-Segel model with log-concentration input given by Eq. \eqref{eq:logsensing}, $D = 865 \, \upmu \text{m}^2$/s, $\alpha = 1850 \, \upmu \text{m}$, $v_{\text{drift}} = 5.2 \, \upmu$m/s.}
\label{fig:densityks}
\end{figure*}

Experiments were carried out immediately after plasma etching as the permeability of gases slowly decreases thereafter.
Cultures of \textit{S. rosetta} were introduced at the 
inlet of the device, and both the inlet and outlet were then closed to prevent evaporative flows.
A gradient of oxygen was set up by having air flowing in one of the side channels and nitrogen in the other.

\begin{figure}[b]
\centering
\includegraphics[width=0.4\textwidth]{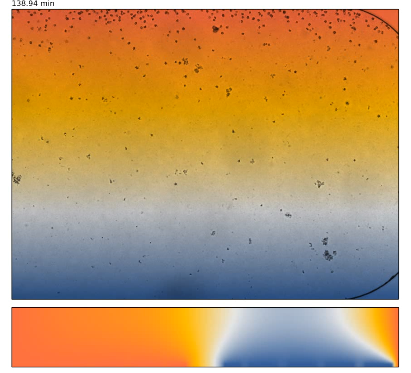}
\caption{\textsc{Video 1}. Experimental videos of aerotaxis (top) and oxygen gas simulation (bottom) [as in Fig. \ref{fig:gassim}]. 
Experimental videos are colored by the output of the gas simulation.
Video 1: Colonies migrating from one side to other after a swap of nitrogen and air [138 -- 148 min. in experiment of Fig. \ref{fig:densityks}].
Video 2: Colonies migrating from the middle to the sides after a change from nitrogen only in the two gas channels to air only [45 -- 55 min. in experiment of Fig. \ref{fig:densityks}]. \textit{(\small ''Video\_1\_swap.avi'' and ''Video\_2\_stay.avi'')}.}
\label{fig:migrationvideos}
\end{figure}

\subsection{Aerotaxis in Choanoflagellates}
Our main experimental result, shown in 
Figs. \ref{fig:densityks}a,b, is the observation that \textit{S. rosetta} colonies accumulate at the oxygen-rich side and away from the oxygen-poor side,
\textit{i.e.} that they are aerotactic.

This result has two main implications for choanoflagellates: first, they must have an oxygen-sensing mechanism. Given their evolutionary proximity to animals in the tree of life, an interesting question is naturally the similarity to the oxygen-sensing mechanism of animals.
Secondly, choanoflagellates must have a mechanism to bias their swimming in order to swim towards oxygen-rich neighborhoods.
We also found aerotaxis in the unicellular fast swimmer form \cite{Dayel2011} of \textit{S. rosetta} (see SI), showing that this is not an exclusive phenomenon to colonies.

With the present microfluidic device we can explore more details of choanoflagellate aerotaxis by dynamically changing the 
oxygen boundary conditions, for instance by flipping the gradient direction or by removing all oxygen influx after a 
uniform distribution has been reached.
Fig. \ref{fig:densityks}c shows the result of such a dynamic experiment over the course of $\sim$ 3.5 hours.
The density is normalized for each frame resulting in some noise due to missing colonies in the tracking.
Many repetitions of the experiment show that the behavior in Fig. \ref{fig:densityks}c is highly repeatable and robust to changes in 
the details of the cycling protocol (See Fig. \ref{fig:densityks} Supplement).  
For consistency the figures in the main text 
are based on this specific protocol.
Whenever one gas channel contains oxygen and the other nitrogen, the colonies swim towards the oxygen-rich side as further shown in Video 1 of Fig. \ref{fig:migrationvideos}.
In the time after a gas channel swap, the slope of the maximum density reveals the ensemble drift velocity $v_{\text{drift}}$.
When there is oxygen in both gas channels, we observe that the density reaches an approximately uniform distribution within the 
time frame of the experiment.
For periods in which nitrogen flows in both channels, this is not the case.
Under these experimental conditions, the colonies accumulate in the middle of the chamber, where there is still some residual oxygen,
as further shown in Video 2 of Fig. \ref{fig:migrationvideos}.
The fact that in this nitrogen-only configuration the colonies accumulate mid-chamber shows that accumulation does not depend on the 
presence of a nearby surface. 
With only nitrogen flowing, eventually there will be no oxygen gradient.
Nonetheless, we observed the colonies to stay in the 
middle of chamber even after 90 minutes (see Fig. \ref{fig:densityks} Supplement).
At that time, the highest oxygen levels are 
estimated by the diffusion equation to be less than $\sim 0.2 \, \%$.
This contrasting behavior between the oxygen-only and nitrogen-only configurations suggests an asymmetry in the aerotactic response, the nature of which can be made precise by studying the population dynamics.

\begin{figure*}[t]
\centering
\includegraphics[width=1\textwidth]{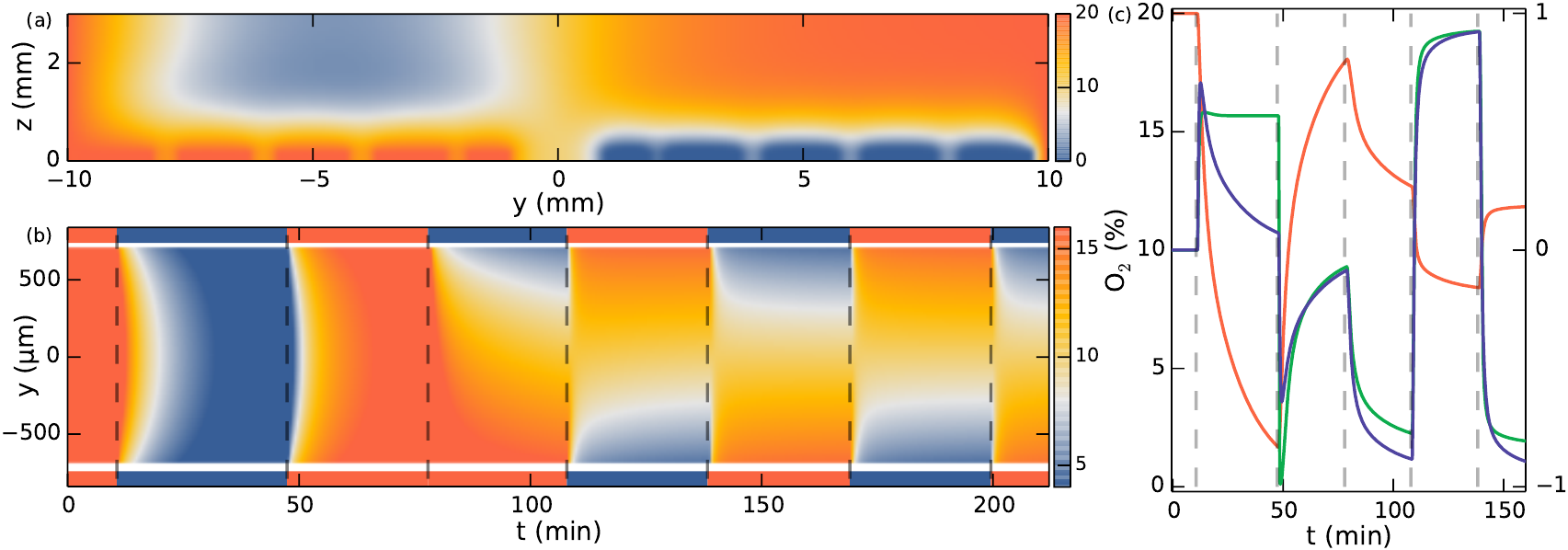}
\caption{Simulation of oxygen concentration in microfluidic device. (a) Simulation of 2D slice of the device. Snapshot shows $t = 110$ min., $\sim$ 1.5 min. after the swap. (b) Evolution of oxygen concentration at $z= 100 \, \upmu$m (c) Simulation at $y = -250 \, \upmu$m. Oxygen percentage in red (left axis), and spatial gradient in purple normalized to fit in $[-1,1]$ (right axis), response function $\tanh( \alpha \nabla c(\x, t) / c(\x, t) )$ in green (right axis).}
\label{fig:gassim}
\end{figure*}

\subsection{Population Dynamics}
The spatio-temporal evolution of the choanoflagellate colony population within the observation chamber is studied here with the
Keller-Segel model \cite{Keller1971}, which has broad applicability for taxis \cite{Tindall2008}.
In describing the phenomenon of aerotaxis the two quantities of interest are the colony population density $\rho(\x,t)$,
and the oxygen concentration $c(\x,t)$.  These evolve according to the coupled equations
\begin{equation}
\frac{\partial \rho}{\partial t} = D\nabla^2 \rho - \bm{\nabla} \cdot \left(\mathcal{F}[c]  \, \rho  \right) 
\label{eq:ksgeneral}
\end{equation}
and
\begin{equation}
\frac{\partial c}{\partial t} = \bm{\nabla} \cdot  \left(D_c \bm{\nabla} c \right)~.
\label{eq:diffusionoxygen}
\end{equation}
Here $D$ and $D_c(\x)$ are the colony and oxygen diffusion constants, the latter of which varies with position inside the
microfluidic device, with values  $D_{c, \text{PDMS}} = 3.55 \times 10^{-3} \, \text{mm}^2/\text{s}$ 
in PDMS \cite{Cox1986} and  $D_{c, \text{water}} = 2.10 \times 10^{-3} \, \text{mm}^2/\text{s}$ 
in water \cite{Cussler1997}. 
The functional $\mathcal{F}$ specifies the population drift velocity's dependency on the local oxygen concentration.  
In a fixed, linear gradient one would have $\mathcal{F}= \mathbf{v}_{\text{drift}}$. 

Using in-house software, we solve Eq. \eqref{eq:diffusionoxygen} on a cross-section of the microfluidic device with time dependent boundary conditions corresponding to the experimental protocols.
Gas channels with oxygen flowing have the condition $c = 20 \, \%$ (for theory the unit of oxygen is not important and we find percentage to be the most intuitive measure). 
For channels with nitrogen flowing $c = 0$. 
Glass interfaces have no-flux conditions $\bf{\hat{n}}\cdot \bm{\nabla} c = 0$, where $\bf{\hat{n}}$ is 
the surface normal.

A snapshot from the numerical studies is shown in Fig. \ref{fig:gassim}a at a time following a swap of nitrogen and oxygen channels, thus showing residual oxygen above the channels now filled with nitrogen and vice versa.
The simulation can now be evaluated at the position of the observation chamber. Note that the no-flux conditions 
at the glass interface render the concentration gradients in the $z$-direction very small, so the precise height of evaluation is
not significant. The simulation with boundary conditions corresponding to those in the experiment of Fig. \ref{fig:densityks}c is shown in Fig. \ref{fig:gassim}b.
To a very good approximation the concentration field is constant along the $x$-direction.
Fig. \ref{fig:gassim}c shows the oxygen field evaluated at $y = -250 \, \upmu$m (red curve).
Neglecting the consumption of oxygen by the colonies, these results provide the input concentration field $c$ for
the Keller-Segel model.

\begin{figure*}[t]
\centering
\includegraphics{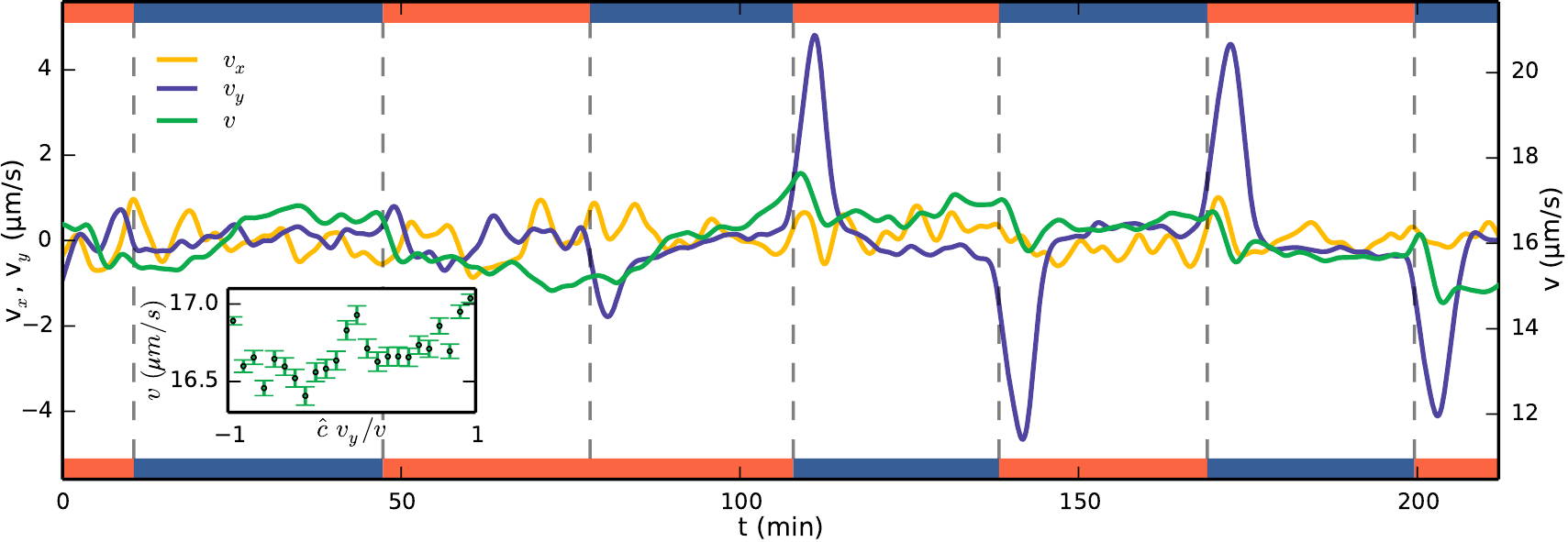}
\caption{Running mean velocity statistics, showing that the primary mechanism of aerotaxis is not by modulation of swimming speed.
Evolution of mean speed (green, right axis) and velocity in the $x$-direction (yellow, left axis) and $y$-direction (purple, left axis), $y$ being along the gradient of oxygen. Left and right axes have equal ranges. Side bars indicate gas flowing, oxygen (red) or nitrogen (blue). The peaks of $v_y$ do not quite reach the true drift velocity due to smoothing of the curves. Inset shows the speed as function of alignment with the gas gradient $\hat{c} \, v_y/v$ at times after a swap. $\hat{c} = 1$ if the gradient is up and $=-1$ if down. \label{fig:speed}}
\end{figure*}

The simplest and widely-used response functional in the Keller-Segel model is linear in spatial gradients, 
$\mathcal{F}[c](\x,t) = \beta\bm{\nabla} c(\x, t)$, where $\beta$ is termed the \textit{taxis coefficient}.
Such a response can, however, reach unrealistic drift velocities,\textit{ i.e.} higher than the swimming velocity, if the oxygen 
gradients are large.
This defect can be eliminated by various functional forms \cite{Tindall2008}, such as  the choice
\begin{equation}
\mathcal{F}[c]  = v_{\text{drift}} \tanh \left( \alpha \left| \bm{\nabla} c \right| \right) 
\frac{\bm{\nabla} c}{| \bm{\nabla} c |},
\end{equation}
where $v_{\text{drift}}$ must be smaller than the swimming velocity. This choice behaves linearly for small gradients, 
but tends asymptotically to a maximum value for large gradients.
Most of the behavior of Fig. \ref{fig:densityks}c can be explained by this model (See Fig. \ref{fig:densityks} Supplement), but not the section where nitrogen is flowing in both side channels.
The reason is the aforementioned asymmetry between the sensing-signalling in low versus high oxygen concentrations.
Many biological sensors exhibit logarithmic sensitivity to gradients.
This is the case for the human senses such as hearing and sight, described by the Weber-Fechner law \cite{E.H.Weber1996}, 
and it is also found in certain bacteria \cite{Kalinin2009}.
We propose that logarithmic sensing is a natural explanation for the discussed asymmetry. 
To examine this hypothesis, we note that $\bm{\nabla} \log c = \bm{\nabla} c/c$ and consider
\begin{equation}
\mathcal{F}[c]  = v_{\text{drift}} \tanh \left( \alpha \left | \frac{\bm{\nabla} c}{c} \right| \right) 
\frac{\bm{\nabla} c}{| \bm{\nabla} c |}.
\label{eq:logsensing}
\end{equation}
For unidirectional gradients (say, in the $y$-direction) this expression reduces to $\mathcal{F}[c]  = 
v_{\text{drift}} \tanh(\alpha \, c_y/c) \, \hat{\bf y}$, where $c_y=\partial c/\partial y$.
Since the logarithmic sensing cannot be maintained at infinitely small concentrations, there must naturally 
be some lower cutoff to this expression in absolute concentration levels. 
Nonetheless, from a modelling perspective we can ignore this for the present experiments with nitrogen-only sections lasting less than 1.5 hours 
as discussed. Fig. \ref{fig:gassim}c shows that the spatial oxygen gradient $c_y$ (purple curve) compares to 
$\hat{\bf y} \cdot \mathcal{F}[c]$ of Eq. \eqref{eq:logsensing} (green curve) at all times except in the nitrogen-only section, where the log-response function does not decay towards zero.
A positive value of the response function means a positive ($y$) drift velocity.

The Keller-Segel equation with log-sensing is able to explain all sections of the experiment,
as demonstrated in Figure \ref{fig:densityks}d.
The parameters obtained from a numerical fit include the drift velocity $v_{\text{drift}} = 5.2 \, \upmu$m/s (which should be compared to the ensemble average speed $v = 16.5 \, \upmu \text{m} / \text{s}$) and the  diffusion constant $D = 865 \, \upmu \text{m}^2$/s.  
Using the ensemble-averaged speed, the diffusion constant can be related to 
an effective rotational diffusion constant $D_r = v^2 / 2 D = 0.16 \, \text{s}^{-1}$.

\subsection{Navigation Strategy}
The parameter $v_{\text{drift}}$ represents the coupling between the oxygen-gradient response and the resulting 
population drift, but it is a purely phenomenological quantity in which the underlying microscopic mechanism of 
aerotaxis is hidden.
Strategies of taxis can be categorized into two main classes: \textit{deterministic} and \textit{stochastic}.
In both strategies the swimming organism measures the attractant gradient (for small organisms by some temporal filter \cite{Block1982, Celani2010}). 
A deterministic strategy, then, is one in which the organism directly steers towards the attractant, such as seen in sperm cells that modulate their flagellar 
beat to adjust directly the curvature and torsion of its swimming path in the gradient direction \cite{Friedrich2007, Jikeli2015}. 
Contrasting is a stochastic strategy such as bacterial run-and-tumble locomotion \cite{Berg1993}, where modulation of the frequency of random reorientations biases the motion in the gradient direction without directly steering towards it.

\begin{figure*}[t]
\centering
\includegraphics[width=1.\textwidth]{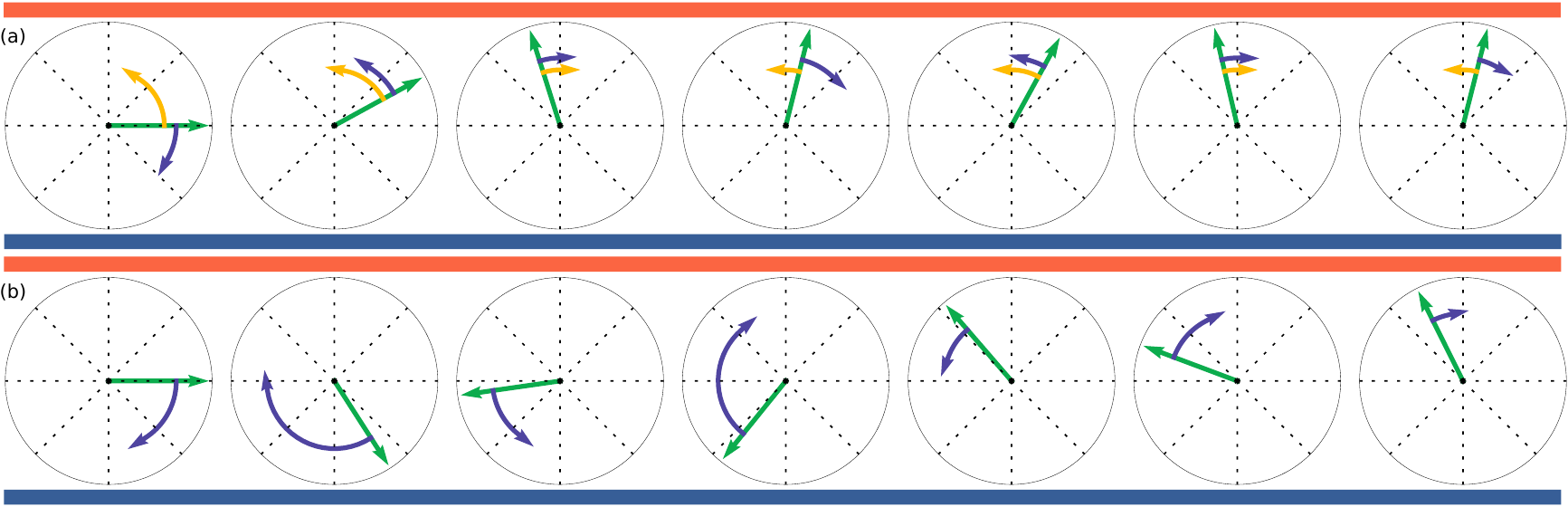}
\caption{Illustration of deterministic and stochastic strategies based on discretised simulations with exaggerated steps. Time evolves from left to right. Orientations, shown by green arrows, are trying to align to up-motion, $\theta=\pi/2$, indicated by red (oxygen) at the top and blue (nitrogen) at the bottom. (a): Deterministic strategy, described by Eq. \eqref{eq:deterministicstrategy}. Deterministic part in yellow and stochastic part in purple. The deterministic part is always in the correct direction. (b): Stochastic strategies, described by Eq. \eqref{eq:stochasticstrategy}. All steps are stochastic, but largest when furthest away from $\theta=\pi/2$. \label{fig:detstoch}}
\end{figure*}

One simple method of taxis results from an organism swimming faster when it is moving up the 
gradient, creating an overall bias towards the attractant. 
With the detailed colony-tracking in the present study it is
possible to test whether this mechanism is in operation with \textit{S. rosetta}.
Figure \ref{fig:speed} shows the evolution of the mean colony swimming speed $v$ (green) as well as the component velocities $v_x$ (yellow) and $v_y$ (purple). 
For most times, the component velocities average to zero, but after a gas channel swap the $y$-component peaks.
The ensemble average swimming speed in these sections, however, does not show an increase, suggesting that a velocity modulation is not the method of taxis.
To quantify this further, the inset of Fig. \ref{fig:speed} shows the swimming speed in these sections plotted against the alignment to the gradient $\hat{c} \, v_y/v$ where $\hat{c} = \pm 1$ signifies the direction of the gradient.
The plot shows a very small ($\sim 3 \%$) change in swimming speed going up the gradient.
Velocity-biased taxis can be described by $\v(t) = v(\hat{p}) \, \hat{p}$,
where \textit{e.g.} $v(\hat{p}) = v \, (1 + \gamma \, \hat{p} \cdot \nabla c / |\nabla c|)$, $\gamma$ being the velocity-modulation taxis parameter.
$\hat{p}$, the direction of swimming, is unbiased by the attractant field $c$ and evolves by rotational diffusion.
To obtain a drift velocity $\sim 1/3$ of the swimming velocity, as found for \textit{S. rosetta} in the previous section, the velocity modulation would have to be $\gamma = 2/3$ for a two-dimensional swimmer and $\gamma=1$ in three dimensions, much larger than the $\sim 3 \%$ observed.
We conclude that the primary mechanism of aerotaxis in \textit{S. rosetta} is therefore not a modulation of swimming speed.

\textit{S. rosetta} colonies swim along noisy helical paths, and each colony displays distinct helix parameters \cite{Kirkegaard2016}. 
To perform any kind of statistical angle analysis, we consider ensemble average quantities: the speed $v$ and 
rotational diffusion $d_r$, and assume that helix rotations average out. 
To distinguish deterministic and stochastic strategies we introduce two effective models and in the following
consider them in the context of a constant oxygen gradient along the $y$-axis, but the generalization is immediate. 
We furthermore ignore translational diffusion due to thermal noise, since this contribution is orders of magnitude smaller than that of active diffusion.
Thus in a quasi-2D system, an organism's path is described by
\begin{equation}
\d \x = \begin{pmatrix}
\cos \theta(t) \\
\sin \theta(t)
\end{pmatrix} v \, \d t,
\label{eq:dxvt}
\end{equation}
where $\theta$ is the instantaneous swimming direction, and we choose motion along the positive $y$-axis 
($\theta = \pi/2$) to be toward the attractant.

In the deterministic model, the organisms actively steer towards the gradient. We model this with the Langevin equation
\begin{equation}
\d \theta = \epsilon_d \cos \theta \, \d t + \sqrt{2 d_r} \, \d W(t),
\label{eq:deterministicstrategy}
\end{equation}
where $W(t)$ is a Wiener process for which $\langle \d W(t) \, \d W(t') \rangle = \delta(t-t')$. 
This process is illustrated in Fig. \ref{fig:detstoch}a, yellow arrows showing $\epsilon_d \cos \theta \, \Delta t$ and purple arrows $\sqrt{2 d_r} \, \Delta W$.
For the stochastic model we take a `continuous' version of run-and-tumble, in which the rotational diffusion is modulated
\begin{equation}
\d \theta = \sqrt{2 d_r (1 - \epsilon_s \sin \theta )} \, \d W(t),
\label{eq:stochasticstrategy}
\end{equation}
where $-1 \leq \epsilon_s \leq 1$ and the multiplicative noise is interpreted in the It\={o} sense.
This process is illustrated in Fig. \ref{fig:detstoch}b.
In both models, $\epsilon$ is an effective ensemble average response.
To compare to the experiments, $\epsilon$ should be replaced by a function coupled to the gas concentration field through $\mathcal{F}[c]$, thus coupling to Eq. \eqref{eq:dxvt}. 
The steady state gradient in the experiment is approximately linear, and thus as a first approximation Eqs. (\ref{eq:deterministicstrategy}, \ref{eq:stochasticstrategy}) should describe the angle statistics in the time between the swap of gas channels and reaching the opposite side. 

\begin{figure}[tb]
\centering
\includegraphics{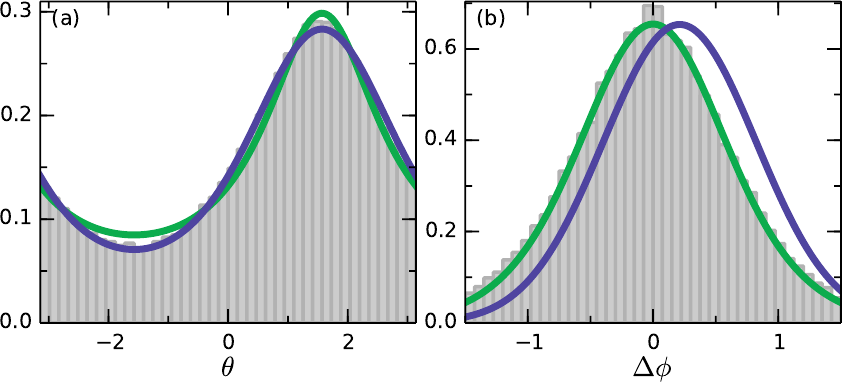}
\caption{Angle statistics. Deterministic model in green and stochastic in purple. (a) Distribution of $\theta$. $\theta = \pi / 2$ is along the gradient. (b) Change in angle $\Delta \phi$ for $\Delta t = 0.65$ s. Positive change corresponds to a turn towards the gradient. Deterministic parameters: $\epsilon_d = 0.28 \, \text{s}^{-1} , \, d_r = 0.52 \, \text{s}^{-1}$. Stochastic parameters: $\epsilon_s = 0.55, \, d_r = 0.33\, \text{s}^{-1}$. \label{fig:reorient}}
\end{figure}

In our experiments, after a flip in the oxygen gradient direction, the colonies reach the opposite wall in a time comparable to
that for oxygen to diffuse across the chamber.  Thus, a true steady state is not reached, but during the intermediate times a steady
state approximation is accurate. 
Solving the Fokker-Planck equations corresponding to the systems Eq. \eqref{eq:deterministicstrategy} and \eqref{eq:stochasticstrategy} in steady-state, we obtain theoretical angle distributions (see Methods).
Figure \ref{fig:reorient}a shows the angular distribution data, where, for the purposes of
displaying all the data in a single graph, we have let 
$\theta \rightarrow - \theta$ for times when the oxygen gradient were pointing down. 
Both models are able to describe the data well, with the deterministic model (purple) fitting the down-gradient swimming best, 
and the stochastic model (green) fitting the up-gradient swimming best.
Both fits involve a single parameter, $k_d = \epsilon_d/\sqrt{2d_r}$  in the deterministic model and $\epsilon_s$ in the stochastic one.

We now move beyond steady-state distributions to examine the detailed statistics of the trajectories themselves,
and define the angle turned by a colony in a time $\Delta t$ as $\Delta \phi = |\theta(t+\Delta t) - \pi/2| - |\theta(t) - \pi/2|$ 
such that it is positive if the turn is in the direction of the gradient and negative otherwise, and choose $\Delta t$ low enough 
that $-\pi < \Delta \phi < \pi$.
The equations \eqref{eq:deterministicstrategy} and \eqref{eq:stochasticstrategy} imply distributions of $\Delta \phi$ (see Methods).
Figure \ref{fig:reorient}b shows the best fit of both models to the data. For the stochastic model (green) $\epsilon_s$ is known and 
the fit is in $d_r$ and matches well. The deterministic (purple) is constrained by $k_d = \epsilon_d / \sqrt{2 d_r}$, and the fit can be 
done in $d_r$ as well. We see clearly that the deterministic model does not provide a satisfactory fit to the data. 
In detail, the value of $\epsilon_d$ needed to fit the data in Fig. \ref{fig:reorient}a shifts the mean of $p_d(\Delta \phi)$ in the 
positive direction.
This result persists with any amount of smoothing applied to the data, averaging out active rotations.
We thus conclude that the colonies navigate by a stochastic strategy, and that the ensemble angle statistics can be captured 
by this simple model.

The population dynamics must be consistent with the above strategy model.
Having shown that the data favor a stochastic model,
we may now couple Eq. \eqref{eq:stochasticstrategy} to Eq. \eqref{eq:dxvt} 
and let the Fokker-Planck equation (Methods) replace the Keller-Segel model. 
Such an approach leads to similar results as Fig. \ref{fig:densityks}d. 
Furthermore, we now recognize that the Keller-Segel model is a quasi-stationary approximation and we can calculate the stationary-approximation drift velocity
\begin{equation}
v_{\text{drift}} = v \, \langle \sin \theta \rangle = v \left[ 1/\epsilon_s - \sqrt{1/{\epsilon_s}^2 - 1} \right].
\end{equation}
In other words $\epsilon_s = (2 \, v \, v_{\text{drift}})/(v^2+v_{\text{drift}}^2)$ is the ratio of the squared geometric mean to the quadratic mean of the average and drift velocities. 
For the fitted $\epsilon_s = 0.55$, $v_{\text{drift}} \approx 0.3 \, v \approx 5 \, \upmu \text{m} / \text{s}$, consistent with the fitted value in the Keller-Segel model.

\section{Discussion}
The experimental observation that choanoflagellates are aerotactic raises of number of questions. One concerns the actual sensing mechanism.
In certain aerotactic bacteria, it is known that oxygen is sensed not directly, but indirectly via energy-sensing \cite{Taylor1999}.
In animals, oxygen levels can be sensed by the hypoxia-inducible transcription factor pathway, which is a highly conserved pathway 
extending all the way back to \textit{Trichoplax adhaerens} \cite{Loenarz2011}.
The question of which mechanism allows aerotaxis in choanoflagellates reimains open.

We have shown that in spatio-temporal varying environments, consideration of population dynamics can distinguish between linear- and 
logarithmic-sensing mechanisms. 
The fact that such conclusions can be made from 
population dynamics alone allows similar analysis to be done in dense populations, where long individual tracking is not possible.
This is a clear advantage of applying models of the Keller-Segel type directly to experimental data.
By logarithmic sensing, what is meant here is that there is a logarithmic component somewhere in the sensing-signalling cascade.
It could be at the sensing, signalling, or response stage.
In any case, logarithmic sensing is a key attribute for survival: it allows sensing of and reacting to gradients in very low concentration environments, 
while still being able to effectively navigate along large gradients.

We have described the stochastic strategy of \textit{S. rosetta} in terms of an effective model. The effective bias 
parameter $\epsilon_s$ is the result of flagella modulation.
Flagella can be imaged on colonies stuck to the microscope slide \cite{Kirkegaard2016}, but
measuring directly the flagella modulation is infeasible, since the oxygen changes that can be induced in a microfluidic device 
are on a time scale of minutes, whereas the flagella beating is on a time scale of tens of milliseconds \cite{Kirkegaard2016}. 
The swimming trajectories suggest that this modulation is a mixture of many types, ranging from slow to vigorous, as also seen in other organisms \cite{Jikeli2015}, but direct imaging is needed to make more quantitative statements.

Finally, we note that sponges, the earliest animal and the closest animal relative to choanoflagellates, were recently shown to have 
very low oxygen requirements, disputing the claim of oxygen as the trigger of animal evolution \cite{Mills2014, Nursall1959}.
Our results add to this discussion, in showing that their evolutionary precursors, the choanoflagellates, have evolved to navigate actively towards oxygen.

\section{Materials and Methods}
\subsection{Culturing \textit{S. rosetta}}
\textit{S. rosetta}  were cultured polyxenicly in artificial seawater ($36.5$ g/L Marin Salts (Tropic Marin, Germany)) with organic 
enrichment ($4$ g/L Proteose Peptone (Sigma-Aldrich, USA), 
$0.8$ g/L Yeast Extract 
(Fluka Biochemika)) at $15$ $\upmu$l/mL, and grown at 23$^{\circ}$C, split weekly.
Cultures were centrifuged to reach the high concentrations, $\sim 5 \times 10^6 \, \text{ml}^{-1}$, used in experiments.

\subsection{Microfluidic Device}
Microfluidic devices were manufactured using standard soft-lithography techniques. The master was produced by spinning SU8-2075 (MicroChem, USA) at 1200 rpm to a thickness of $\sim 115 \, \upmu$m. Chambers were cast in PDMS, Sylgard 184 (Dow Corning, USA), and plasma etched to the glass slides. Cultures were concentrated by centrifugation before loaded into the device and two gas cylinders were connected via a system of valves to the gas channels of the device.
Experiments were filmed in bright field at 10 fps.

\subsection{Data processing}
Colonies were tracked using custom in-house software. Density distributions were estimated by first calculating
\begin{equation}
\eta(y) = \frac{\sum_i \exp(-(y_i-y)^2/2\sigma^2)}{\int_{y_0}^{y_1} \exp(-(y'-y)^2/2\sigma^2)  \, \d y'},
\end{equation}
where $y_i$ are the tracked positions, $\sigma$ a standard deviation of separation, and the denominator adjusts for boundary effects. Hereafter $\rho(y) = {\eta(y)}/{\int_{y_0}^{y_1} \eta(y') \, \d y'}$ is the normalized density. For velocity and angle statistics, the tracked positions were linked by proximity.

Oxygen diffusion and Keller-Segel equations were numerically solved using in-house software with finite difference spatial discretisation and implicit time-stepping.

\subsection{Effective stochastic models}
Given the stochastic dynamics \eqref{eq:deterministicstrategy} for individual particles following a deterministic strategy, the probability distribution
function $p_d(\theta,t)$ for the population obeys the Fokker-Planck equation
\begin{equation}
\frac{\partial p_d}{\partial t} = d_r \frac{\partial^2 p_d}{\partial \theta^2}
-\epsilon_d \frac{\partial}{\partial \theta} \left( \cos (\theta) \, p_d \right).
\end{equation}
The steady state distribution is found to be a von-Mises distribution
\begin{equation}
p_d(\theta) = \frac{1}{2 \pi \, I_0(\epsilon_d/\sqrt{2 d_r})} \exp \left( \frac{\epsilon_d \sin \theta}{\sqrt{2 d_r}} \right),
\end{equation}
where $I_0$ is the modified Bessel function of order zero. On the other hand, for the Fokker-Planck equation for the stochastic model,
\begin{equation}
\frac{\partial p_s}{\partial t} = \frac{\partial^2 }{\partial \theta^2} \left(d_r (1 - \epsilon_s \sin \theta) \,  p_s\right),
\end{equation}
we find the steady-state distribution
\begin{equation}
p_s(\theta) = \frac{1}{2 \pi} \frac{\sqrt{1 - {\epsilon_s}^2}}{1 - \epsilon_s \sin \theta}.
\label{eq:stochasticthetastationary}
\end{equation}

In the time right before a channel swap, we have $p(\theta) \approx 1/2 \pi$, since the colonies stay
near the wall. 
In the deterministic model, for small $\Delta t$, $\Delta \phi = |\theta(t+\Delta t) - \pi/2| - |\theta(t) - \pi/2|$  is composed of deterministic 
$\delta = \epsilon_d \cos(\theta) \Delta t$ and stochastic $\xi = \sqrt{2 d_r} \Delta W$, where $\Delta W \sim \sqrt{\Delta t}$. 
Since we are assuming $p(\theta) = 1/2 \pi$, we have $p(\delta) = 1/\pi \sqrt{(\epsilon_d \Delta t)^2 - \delta^2}$ for $\delta \in [-\epsilon_d \Delta t; \, \epsilon_d \Delta t]$. The distribution of $\Delta \phi = |\delta| + \xi$ is then found as the convolution
\begin{equation}
p_d(\Delta \phi) = \int_{0}^{\epsilon_d \Delta t} \frac{\exp(-(\Delta \phi - \delta)^2/4 \d_r \Delta t)}{\sqrt{\pi^3 d_r \Delta t \, [ (\epsilon_d \Delta t)^2 - \delta^2 ] }} \, \d \delta,
\end{equation}
which can be evaluated numerically by Gaussian quadrature. In the stochastic model there is only $\xi = \penalty 0 \sqrt{2 d_r(1-\epsilon_s \sin \theta)} \Delta W$, but this is conditional on $\theta$. We marginalize for the final distribution
\begin{equation}
p_s(\Delta \phi) = \int_{-\pi}^{\pi} \frac{\exp(-\Delta \phi^2 / 4 d_r(1- \epsilon_s \sin \theta) \Delta t )}{\sqrt{ 16 \pi^3 d_r(1- \epsilon_s \sin \theta) \Delta t } } \d \theta.
\end{equation}

\subsection{Acknowledgements}
Work supported by the EPSRC and St John's College (JBK), ERC Advanced Investigator Grant 247333 
and a Wellcome Trust Senior Investigator Award.

\setcounter{figure}{2}
\begin{figure*}[tb]
\centering
\includegraphics{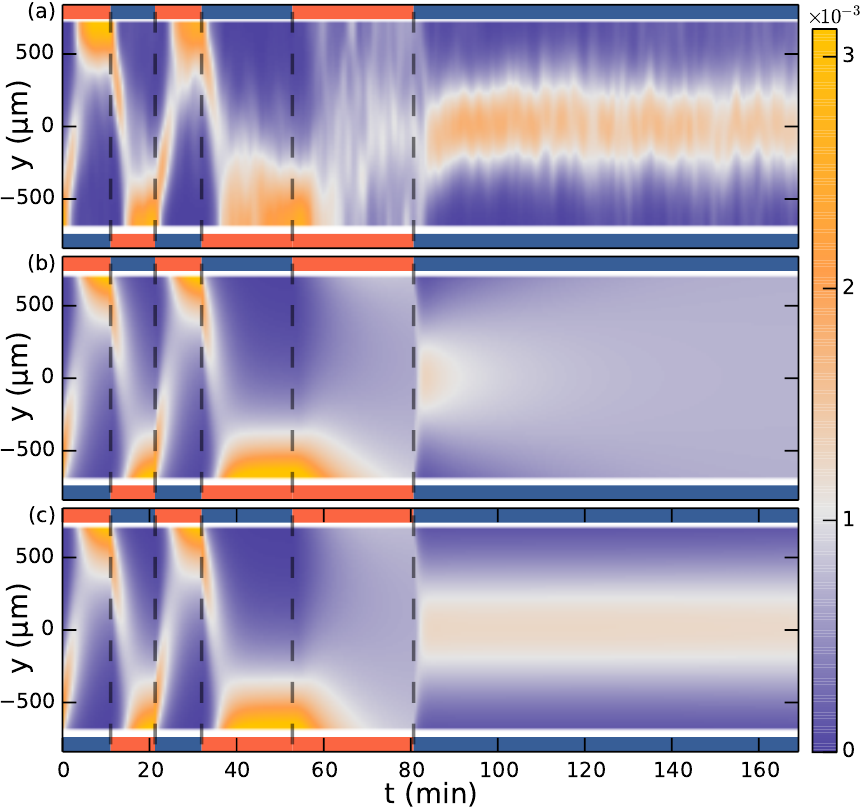} 
\caption{\textsc{Supplement.} Seperate aerotaxis experiment. (a) Density evolution of \textit{S. rosetta} during experiment [colorbar units in $\upmu\text{m}^{-1}$]. $N_{\text{colonies}} \sim 60$, concentration $\sim 2 \cdot 10^6 \, \text{mL}^{-1}$. (b) Keller-Segel model with linear concentration input $\mathcal{F}[c]  = v_{\text{drift}} \tanh(\alpha \, c_y)$, with $\alpha = 8.2$, and otherwise same parameters as in main text Fig. \ref{fig:densityks}. (c) Keller-Segel model with log concentration input, same parameters as in main text Fig. \ref{fig:densityks}. Experiment and simulation was started in steady-state configuration of oxygen down and nitrogen up.}
\label{fig:densityks_SI}
\end{figure*}

\end{document}